\documentclass[useAMS,usenatbib,referee]{mn2e}

\usepackage{epsfig}

\usepackage{graphicx}

\title[A model for the spectrum of the inner jets of Cen A]
  {A model for the electromagnetic spectrum of the inner jets of Cen A}

\author[M. Orellana \& G.E. Romero]
  {M. Orellana $^{1,2}$\thanks{Fellow of CONICET, email: morellana@fcaglp.unlp.edu.ar}
  , G.E. Romero$^{1,2}$\thanks{Member of CONICET}\\
  $^1$Instituto Argentino de Radioastronom\'{\i}a (CCT-La Plata, CONICET), Bs.As., Argentina\\
  $^2$Facultad de Ciencias Astron\'omicas y Geof\'{\i}sicas - UNLP, Bs. As., Argentina}

\date{Released 2008, December 4th}

\pagerange{\pageref{firstpage}--\pageref{lastpage}} \pubyear{2002}

\def\LaTeX{L\kern-.36em\raise.3ex\hbox{a}\kern-.15em

    T\kern-.1667em\lower.7ex\hbox{E}\kern-.125emX}

\begin{document}

\label{firstpage}

\maketitle

\begin{abstract}

Centaurus A, the closest active galaxy, has been detected from radio to high-energy gamma-rays. The synchrotron radiation by extremely high energy protons may be a suitable mechanism to explain the MeV to GeV emission detected by the instruments of the Compton Gamma-Ray Observatory, as coming from the inner jets. This scenario requires a relatively large magnetic field of about $10^4$ G that could be present only close to the central black hole. We investigate the spectral energy distribution (SED) resulting from a {\em one-zone} compact acceleration region, where both leptonic and hadronic relativistic populations arise.

We present here results of such a model, where we have considered synchrotron radiation by primary electrons and protons, inverse Compton interactions, and gamma-ray emission originated by the inelastic hadronic interactions between relativistic protons and cold nuclei within the jets themselves. Photo-meson production by relativistic hadrons were also taken into account, as well as the effects of secondary particles injected by all interactions. The internal and external absorption of gamma rays is shown to be of great relevance to shape the observable SED, which was also recently constrained by the results of HESS.

\end{abstract}

\begin{keywords}

radiation mechanism: non-thermal -- galaxies:active -- galaxies: individual: Centaurus A.

\end{keywords}

\section{Introduction}

The elliptical galaxy NCG 5128 is the stellar body of the giant double radio source Centaurus A (Cen A). The whole radio source extends $\sim 10^\circ$  on the sky, and it can be resolved down to sub-arcsecond scales in the inner radio structures which correspond to several pc because of its proximity ($3.84\pm0.35$ Mpc, Rejkuba 2004). Cen A is one of the best examples of a radio-loud AGN (a Fanaroff-Riley Class I galaxy) viewed from the side ($\sim 70^\circ$) of the jet axis. 
The galaxy has an absorbing band of gas and dust projected across its stellar body. The nucleus ejects linear radio/X-ray jets, becoming sub-relativistic at a few parsec. At about 5 kpc from the core, the jets expand into plumes, and there are huge radio lobes that extend beyond the plumes out to 250 kpc.\\

For further information, the reader is referred to the review by Israel (1998), which focuses on the observed properties of Cen A and their phenomenological interpretation \footnote{Updated references can be found at the dedicated Cen A web page  {\tt http://www.mpe.mpg.de/Cen-A/}.}. Besides of the electromagnetic emission, Cen A has recently called the attention because of a striking clustering of ultra high-energy cosmic ray events observed by the Pierre Auger Observatory around its location (Abraham et al. 2007, 2008). This strongly suggests that protons can be efficiently accelerated up to very high energies in Cen A, as it has been suggested long time ago (Romero et al. 1996).\\

Historically, Cen A has exhibited strong variability (more than one order of magnitude in flux) at X-rays (Bond et al. 1996). Noticeable variations are also present at radio wavelengths, likely powered by accretion events. Very high resolution (VLBI) radio measurements are needed to separate the extremely compact nucleus from its surroundings. The difference between jet and counter-jet brightnesses can be explained by mildy relativistic Doppler beaming, which enhance the radiation of the approaching jet. From the observed brightness ratio, it appears that the northeastern jet is approaching and the southwestern jet receding at moderately relativistic speeds $v$ $\geq$ 0.45c (Jones et al. 1996; Tingay et al. 1998; see also Bao \& Wiita 1997).\\

Cen A has been observed repeatedly by all the instruments of the Compton Gamma-Ray Observatory (CGRO). The spectrum determined by the Energetic Gamma-Ray Experiment Telescope (EGRET, Thompson et al. 1995) and the Compton Telescope (COMPTEL) can be fitted with a broken power law, with a spectral index steepening at high energies. The spectral energy distribution (SED) shows a broad peak around $E_{\rm peak}\sim 0.1$ MeV, with a luminosity of  $L_{\gamma} \sim 3 -5\times 10^{42}$ erg s$^{-1}$ assuming a distance of 3.5 Mpc. At higher energies, only an integral flux upper limit, with a significance of $\sim 0.4\sigma$, was obtained by HESS through a short 4.2 hs live-time exposure (Aharonian et al. 2005). These observations, however, are not simultaneous, something that must be taken into account when pondering any model for the high-energy (HE) emission. \\

Concerning the origin of this HE emission, leptonic models (e.g. an electron-positron beam as part of a two flow model) where the emission is dominated by Inverse Compton (IC), and synchrotron-self Compton mechanisms, have been suggested to explain the nuclear SED of Cen A. These models crucially require of low magnetic fields ($< 10$ G) to allow the electrons to reach high Lorentz factors (e.g. Marcowith et al. 1998, Ghisellini et al. 2005, Lenain et al. 2008).\\

Several authors have proposed that injection of relativistic protons can take place in active galaxies (see Begelman, Rudak \& Sikora 1990, and references therein). Models where the hadrons dominate the HE radiative outcome through photo-meson channels have been applied to blazars (e.g. Mannhein et al. 1991, Mannheim 1993), and hybrid models for extended jet features in AGNs (Aharonian 2002, which includes the proton synchrotron component; see also M$\ddot{\rm u}$cke, \& Protheroe 2001, Reimer et al. 2004, and references therein). The radiating electrons can also be secondary particles produced in inelastic collisions by primary hadrons with ambient nuclei (e.g. Schuster et al. 2001). In that case neutrinos would be emitted along with the gamma rays. \\

Here we have considered the emission of both hadronic and leptonic (primary and secondary) particles in a proton dominated jet, similar to what has been applied recently to low-mass microquasars by Romero \& Vila (2008). We aim to obtain information on the physical parameters related to the nuclear emission of Cen A.\\

\section{Outline of the scenario}\label{outline}

We adopt an Eddington mass accretion rate onto the central supermassive black hole ($M\sim 10^8$ M$_\odot$, Marconi et al. 2000). The accretion disc can be represented by the standard geometrically thin, optically thick, Shakura \& Sunyaev model with a temperature profile
\begin{equation}
T_{\rm disc}(r)=2.2\times 10^5 \dot{M}_{26}^{1/4} r_{14}^{-3/4}\,\,{\rm K},
\end{equation}
where the accretion rate is expressed in units of $10^{26}$ g s$^{-1}$, and the radius in $10^{14}$ cm (Frank, King \& Raine 2002).\\

The inner edge of the disc is given by the last stable orbital radius (for a Schwarzschild black hole $R_{\rm in}=6 R_g$, whereas for a Kerr black hole spinning in the same direction as the disc $R_{\rm in}= R_g$). The disc extends up to $R_{\rm out}= 30R_g$. The gravitational radius of the black hole is $R_g\sim 1.4\times10^{13}$ cm. The hot corona is modeled as a sphere with $R_{\rm cor}=8 R_g$, emitting photons with spectral index $\alpha=1.9$, and with lower and higher cutoffs at 1 eV and 20 keV, respectively. Table~\ref{params} presents the parameters of the model. Note that for the values assumed the Doppler factor is small
\begin{equation}
\delta=\left[\Gamma \left(1-\frac{v}{c}\cos\theta\right)\right]^{-1}\approx 1.\label{Doppler}
\end{equation}

The power redirected to the jets is a fraction of the accreted power, in accordance with the jet/disc symbiosis hypothesis: $L_{\rm j}\sim 0.1 \dot{M}_{\rm acc}$ (e.g. K\"ording et al. 2006). About $\sim 10\%$ of the jet power goes to relativistic particles within a compact region close to the base of the jet, between $z_0=50\,R_g$ and $z_f=5\,z_0$ (e.g. Bosch-Ramon et al. 2006, Romero \& Vila 2008). The kinetic power of relativistic primary particles has two terms, corresponding to the both species originally injected (electrons and protons): $L^{\rm rel}=L_e^{\rm rel}+L_p^{\rm rel}$.

A proton dominated jet is quantified through the ratio $a=L_p^{\rm rel}/L_e^{\rm rel}\approx 100$. The injection distribution follows 
\begin{equation}
Q_{e,p}(E,z)\propto E^{-2} z^{-2}
\label{Qu}
\end{equation} 
for the relativistic particles. Most of the jet content is in the form of a thermal plasma with a mildly relativistic bulk Lorentz factor $\Gamma =3$. This plasma is roughly in equipartition with a tangled magnetic field $B_0\sim 10^4$ G. The jet is assumed to expand in a conical way, thus the density of the cold material within it also decays as $\propto z^{-2}$, and the energetic particles suffer the corresponding adiabatic energy losses.\\

We have obtained the steady state solution of the kinetic equations for the different species of particles taking into account adiabatic and radiative energy losses and the possible escape of the particles from the acceleration region at a timescale $t_{\rm esc}\sim (z_f-z_0)/c$. We neglected diffusion and convection effects in our treatment (see Khangulyan et al. 2007). \\

\begin{figure}
\label{tiempos}
\resizebox{\hsize}{!}{\includegraphics{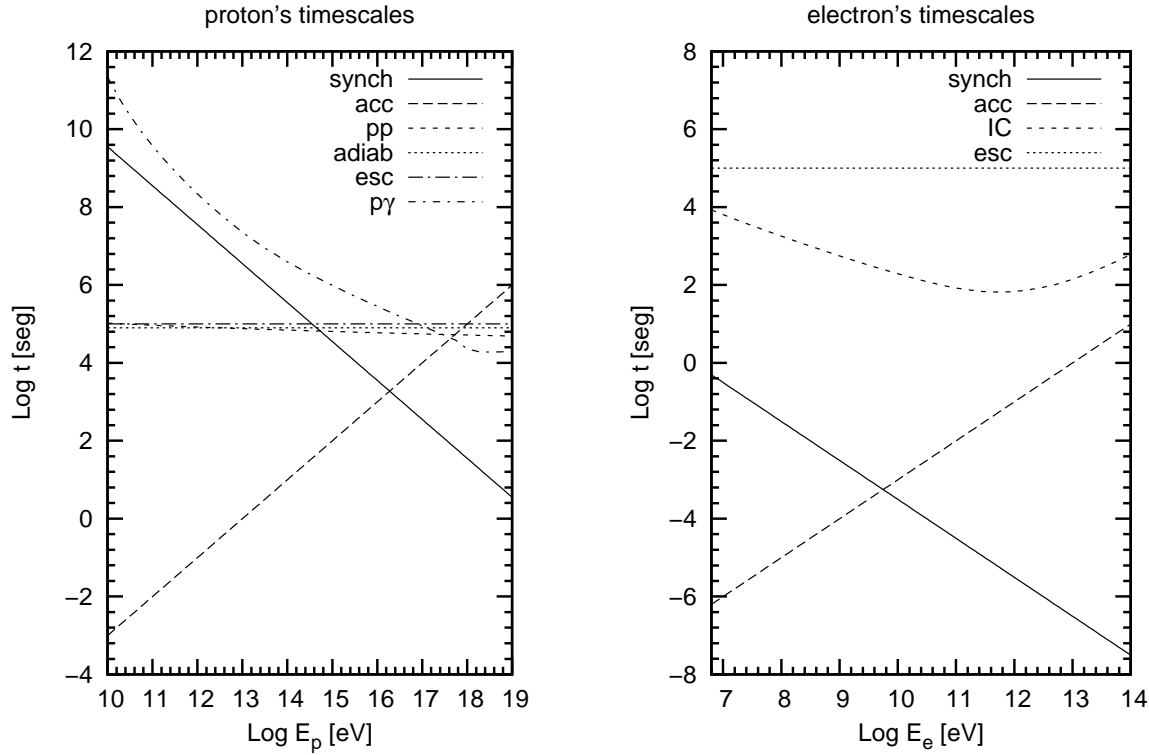}}
\caption{Timescales for energy gain and losses of both protons (left) and electrons (right). The cooling rate by photo-meson production ($p\gamma$) includes only interactions with the proton synchrotron photons (the dominant field).}
\end{figure}

Figure~\ref{tiempos} illustrates the relevant timescales at the base of the jet. The synchrotron energy losses determine the cut-off in the energy distribution of the primary particles, $E_{e,p}^{\rm max}$. We obtain
\begin{equation}
E_{p}^{\rm max}=2.2\times 10^{16}\mbox{ eV, and }\;E_{e}^{\rm max}=4.7\times 10^{9}{\rm eV}.
\label{Emax} 
\end{equation}
At heights $z>z_f$ the relativistic particles cool rapidly.

In expression (\ref{Qu}), the injected relativistic particles have the canonic spectral index for standard first order Fermi diffusive acceleration. We have considered that such acceleration proceeds with an efficiency $\eta$ and on a timescale of $t_{\rm acc}=E/(\eta e c B_0)$. If the steepening in the MeV-GeV spectrum of Cen A is considered as the cut-off signature of proton synchrotron radiation, since the magnetic field is fixed, we obtain a value $\eta\simeq 10^{-4}$ (see Aharonian 2000 for details). The remaining free parameter, the minimum energy of the injected particles, is used to match the MeV luminosity inferred from the CGRO data.  The normalization of the distribution is given by:
\begin{equation}
L_p^{\rm rel}=\int\limits_{\rm Vol}d^3r \int\limits_{E_p^{\rm min}}^{E_p^{\rm max}} dE\,E Q_p(E,z),
\label{norm}
\end{equation}
where the volume Vol is that of the acceleration region. 


\begin{table}

  \caption[]{Values for model parameters.}

 \begin{center}

  \begin{tabular}{lllll}

  \hline\noalign{\smallskip}

 Parameter: description [units] &  values \\[3pt]

   \noalign{\smallskip}\hline


$M$: black hole mass [$M_{\odot}$] & $10^8$ \\

$R_{\rm in}$: inner accretion disc radius [$R_{\rm g}$] & 6\\

$R_{\rm out}$: outer accretion disc radius [$R_{\rm g}$] & 30\\

$R_{\rm cor}$: radius of the corona [$R_{\rm g}$] & 8\\

$z_0$: jet initial point in the compact object RF [$R_{\rm g}$] & 50 \\

$z_f$: end of the acceleration region & 5$z_0$ \\

$\theta$: viewing angle [$^\circ$] & 70\\

$\alpha_{\rm cor}$: corona photon index & 1.9 \\

$\chi$: jet semi-opening angle tangent & 0.1 \\

$\alpha$: relativistic particles power-law index & 2\\

$\Gamma$: macroscopic jet Lorentz factor & 3\\

$a$: relativistic hadron-to-lepton ratio & 100\\

$L^{\rm rel}$: kinetic power in rel. particles [erg s$^{-1}$]& $6\times 10^{43}$\\


$\gamma_{\rm min}$: minimum Lorentz factor of rel. particles & 100\\

$\eta$: acceleration efficiency &$10^{-4}$\\

$B_0$: equipartition magnetic field [G] & $10^4$\\

$n_0$: number density of cold particles [cm$^{-3}$] &$1.5\times 10^{10}$\\

  \noalign{\smallskip}\hline

  \end{tabular}\label{params}

\end{center}

\vspace{-0.2cm}


\medskip{\footnotesize $R_g=G M/c^2$}

\end{table}

\section{The high-energy emission and processing of radiation}\label{emision}

The hadronic synchrotron radiation dominates the SED at MeV-GeV energies. The same primary relativistic protons produce pions through inelastic interactions with the cold material of the jet and with radiation fields. The target photons come from the accretion disc and the corona and are also produced in the jet itself mostly via the proton synchrotron radiation.\\

Inelastic $pp$ collisions produce, through $\pi^0$ decays, $\gamma$ rays in the 1 GeV- 10 TeV range. The computed luminosity is similar to the proton synchrotron one. The primary leptonic synchrotron and IC contributions result of minor relevance.\\ 

The exact calculation of the interaction of the produced radiation fields with the relativistic particles and the radiation self-interaction is a complex problem. As emission is generated in the same volume where it is strongly absorbed or reprocessed, a rigorous formalism should include the solution of a radiative transport equation. Concerning the high-energy photons, along the line of sight $s$, the specific intensity obeys

\begin{equation}\label{radiative}
\frac{dI(E)}{ds}=-I(E,s)\!\!\!\!\!\!\int\limits_{\varepsilon_{\rm min}(E)}^\infty \!\!\!\!\!n_{\rm ph}(\varepsilon,s)\sigma_{\gamma\gamma}(\varepsilon ,E) d\varepsilon
+\! \sum\limits_{\rm proc.} q_i(E,s),
\end{equation}
where the photon emissivities $q_i$ come from the relevant radiative processes at energy $E$, that are usually not the same providing effective absorption, which occurs at energies of the order of $\varepsilon_{\rm min}(E)$.
At the same time, photon annihilation provides an injection term $Q_e$ that enters into the kinetic equation for the distribution of the leptons
\begin{equation}
\frac{\partial}{\partial E}\left(\frac{dE}{dt} n_e(E)\right) + \frac{n_e(E)}{t_{\rm esc}}=\sum Q_e(E),
\label{kinetic}
\end{equation}
where $dE/dt$ includes in-situ acceleration and energy losses (for the general case see, e.g., Ginzburg \& Syrovatskii 1964). The cooling rates and emissivities of the IC and $p\gamma$ processes are also determined by the properties of the radiation fields, therefore the differential equations (\ref{radiative}) and (\ref{kinetic}) are coupled.
Our current treatment is simplified by considering sub-populations that result from each term of the injection function, and the radiative emissivities one-by-one.\\

Whereas the decay of neutral pions leads to $\gamma$-ray emission, the decay of the charged pions originates a population of secondary electrons which extends to higher energies than the primary leptons (see Levinson 2006 and Orellana et al. 2007). The injected distribution function of these products can be computed using the expressions given by Kelner et al. (2006) and Kelner \& Aharonian (2008). For the secondary electrons the kinetic equation to solve is slightly different from that of the primaries, resulting in a break of the distribution at $E_{e}^{\rm max}$ where the acceleration and cooling rates are balanced, and a high-energy cutoff that is a fraction of $E_p^{\rm max}$.\\

\begin{figure}

\resizebox{\hsize}{!}{\includegraphics{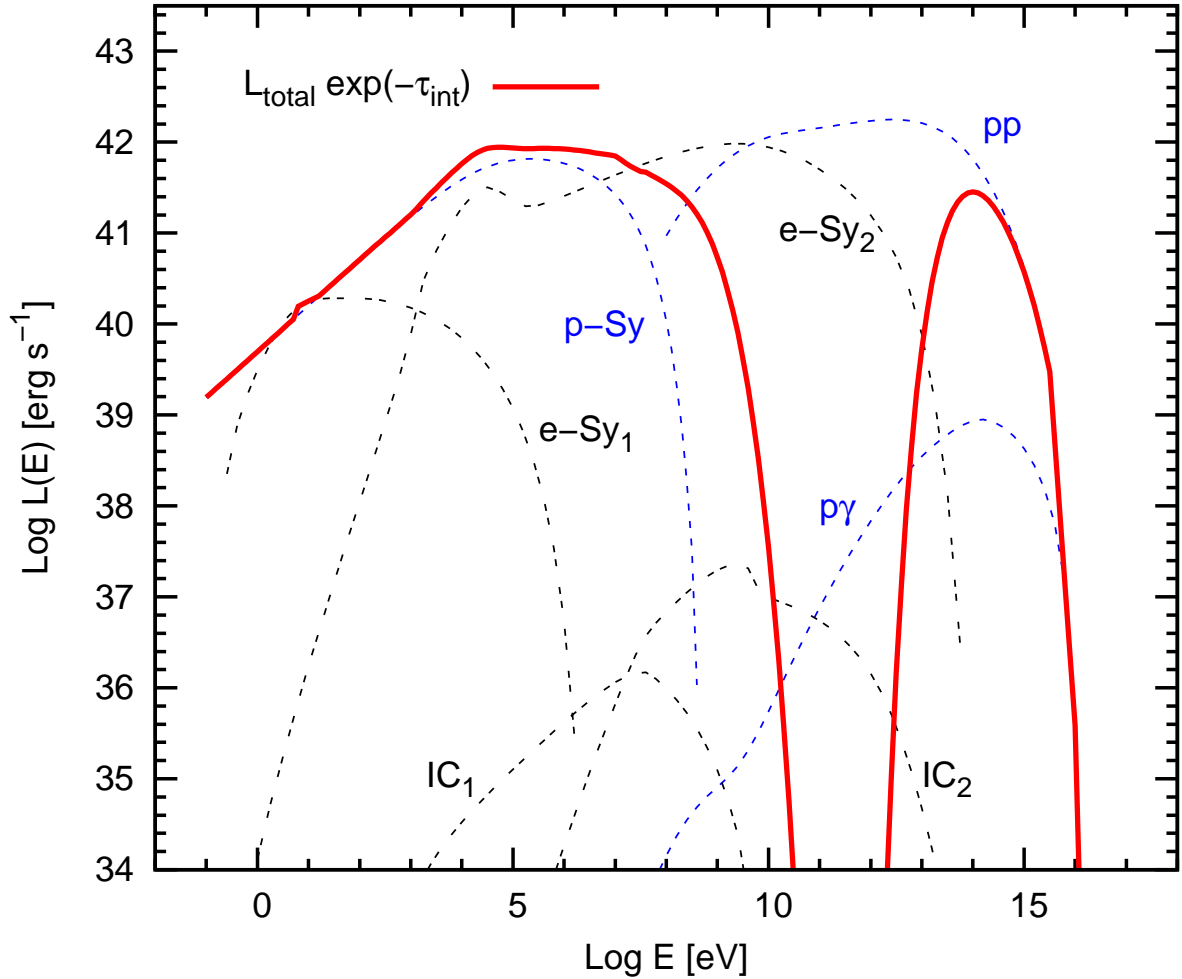}}

  \caption{SED emerging from the inner jets of Cen A, affected only by internal absorption. The contribution from different radiative processes considered are indicated. Sy stands for synchrotron radiation, whereas the subscript 1 or 2 indicates whether the particles are primaries or secondaries.}

\label{sinab}

\end{figure}

In our model the proton synchrotron component extends down to optical wavelengths and provides a dense radiation field which suppress almost the whole $pp$ contribution except for a tail of $\gamma$-ray photons with energies $E\ga 10^{14}$ eV. This internal energy-dependent absorption leads to the injection of relativistic secondary pairs with energies again higher than those of the primary leptons. 
B$\ddot{\rm o}$ttcher \& Schlickeiser (1997) provide useful expressions to calculate the pair injection rates. Figure~\ref{sinab} presents the obtained SED of the inner region of Cen A. 
In the jets, the development of IC electromagnetic cascades is suppressed by the large magnetic field, because the secondaries are mainly cooled by synchrotron radiation (see, e.g., Aharonian, Khangulyan \& Costamante, 2008).\\

\begin{figure}

\centering\includegraphics[width=.4\textwidth]{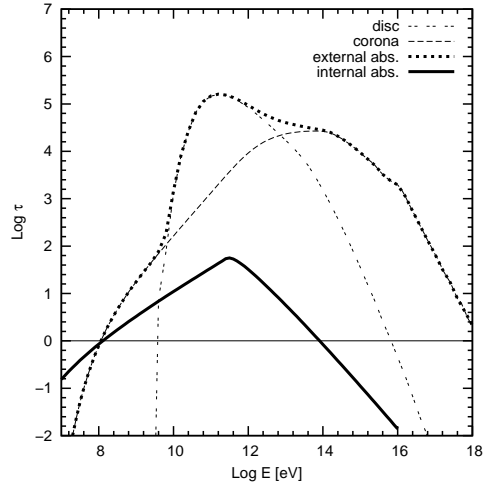}

  \caption{Computed optical depths, for internal absorption in the radiative field dominated by the $p$-synchrotron component, and for external fields taken as the sum of the terms due to the accretion disc and the corona.}\label{taus}

\end{figure}

\begin{figure}

 \resizebox{\hsize}{!}{\includegraphics{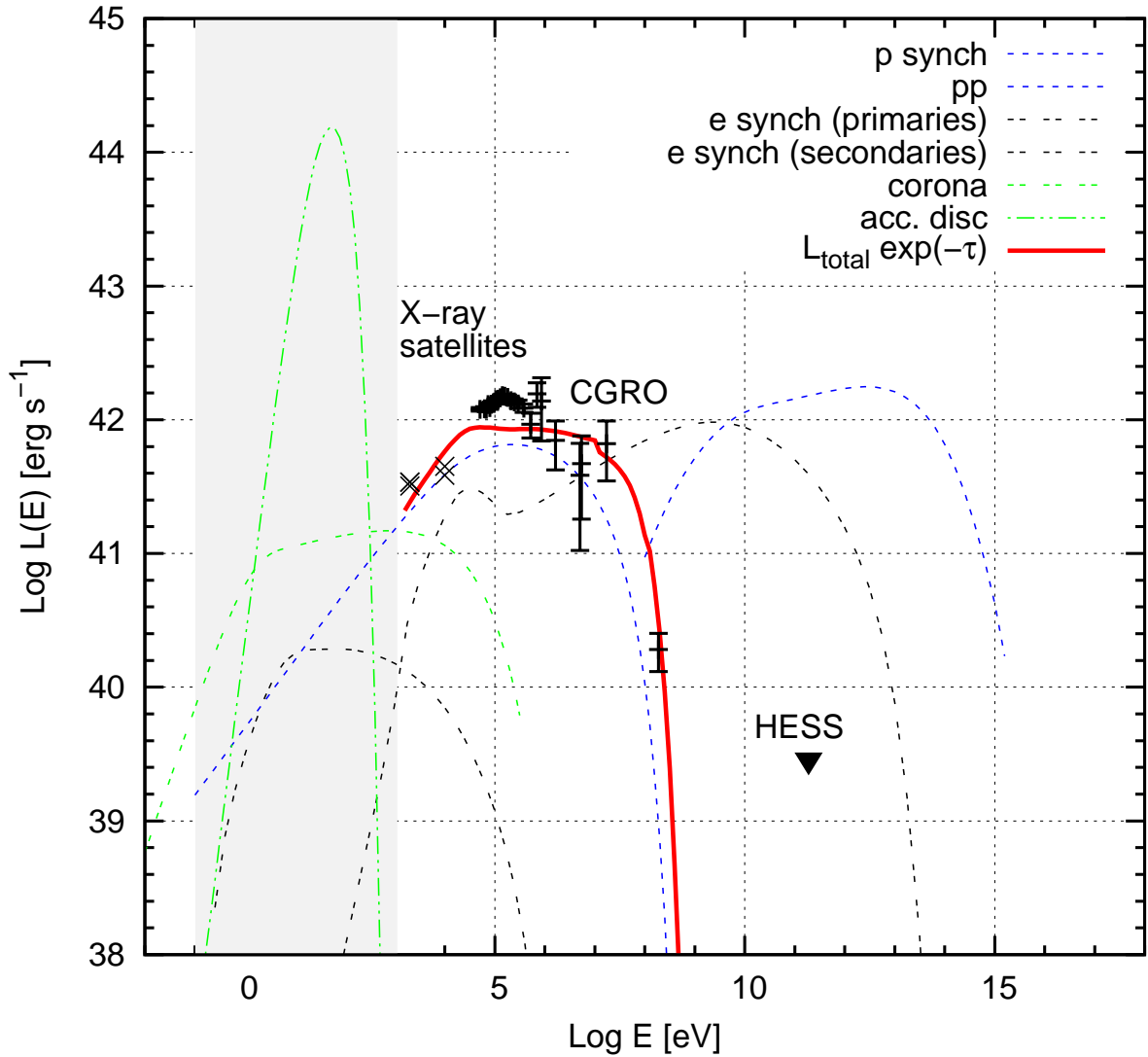}}

  \caption{Computed spectral energy distribution for the inner region of Cen A. The grey band indicates the strong absorption by the dust lane, hiding the AGN core. The high-energy observational constraints are also shown. The data points corresponds to the core of the source.}

\label{sed}

\end{figure}

Outside the jet, in the vicinity of the compact $\gamma$-ray production region, the photon fields provided by the accretion disc and the corona absorb the emerging photons at energies greater than $\sim 100$ MeV. Figure~\ref{taus} shows the dependence of the internal and external optical depths on the energy of the gamma rays. The large viewing angle of the jets (assumed to be perpendicular to the accretion disc)  makes the path of the high-energy photons emerging from the jet to pass close to the disc. Hence geometrical considerations in the calculation of the optical depths can be relevant to shape the detectable spectrum. We have followed the treatment given by Becker \& Kafatos (1995) to compute the opacity provided by the disc, and Dubus (2006) for the corona. Figure \ref{sed} shows the SED corrected by this external absorption plus the internal one. The data points corresponds just to the nuclear emission of the source (Lenain et a. 2008 and references therein). Observational constraints by HESS are also included but, as we mentioned, they were not simultaneously obtained. As Cen A is nearby source, the effect of absorption in the extragalactic background light can be safely neglected. In Figure \ref{sinab} we show the SED that results if only internal absorption is considered. Such a case might be relevant in case of advective inflows onto the black hole.  \\

\section{Discussion}\label{discuss}

The simplified model presented here is capable of reproducing the observed high-energy SED of Centaurus A as coming from a compact acceleration region at the base of its jets. The relativistic content is originally proton dominated but as secondary leptons are copiously produced and the model, as all so-called ``hadronic'' models, is actually lepto/hadronic. We have not considered here the effects of the magnetic field cooling of the decaying pions and muons during their short life-times, but the effects are not expected to be as dramatic as in the case of microquasars (see Reynoso \& Romero 2008 for such a treatment).\\

The low-energy component of the SED (not shown here) that is usually attributed to the core of Cen A (Chiaberge et al. 2001) can result from the radiation of electrons injected in outer regions of the jets, where the magnetic field is lower. Such electrons plus ``fresh'' hadrons could be injected by the decay of neutrons which travel from the acceleration region without suffering synchrotron losses. Such neutrons are generated through the $p+p \rightarrow p+n+\pi^+$ and $p+\gamma\rightarrow n + \pi^{+}$ channels (e.g. Atoyan, 1992). The distance the neutrons travel before suffering $\beta$-decay is given by
\begin{equation}
d_\beta=\gamma_n c \tau_n\simeq\left(\frac{\gamma_n}{10^5}\right)\,{\rm pc}.
\end{equation}
Thus the energy transport by neutral beams to the outer regions of the central source can lead to powerful contributions to the SED at low energies. The electrons injected from neutron decay cool through synchrotron radiation in regions where the field is now low. The electron population will be a power law that will mimic the neutron spectrum, which in turn will reflect the original proton spectrum. The total power produce by these electrons from radio to optical will be $\sim 10^{41-42}$ erg s$^{-1}$, in accordance with the observations, for a magnetic field of $\la 1$ G. Notice that the most energetic neutrons could reach distances of $\sim 100$ pc. They will inject then energetic protons in the inner radio lobes, that could be accelerated then in successive steps up to very high energies (Romero et al. 1996).\\

The physical parameters that were used for tunning our model to reproduce the CGRO observations are the acceleration efficiency $\eta$ and the minimum Lorentz factor of the relativistic particles. The former resulted in a reasonable value of $\eta \sim 10^{-4}$, compatible with acceleration in a parallel shock front (e.g. Protheroe 1998); and the latter is an order of magnitude lower than the one inferred in the SSC model of Chiaberge et al. (2001), namely $\gamma_{\rm min}=2\times 10^3$. \\

Aharonian, Khangulyan \& Costamante (2008) have pointed out that internal absorption could be a problem for the escape of high-energy $\gamma$-radiation from the production region. This absorption can lead to the formation of spectra with almost arbitrary slope in the TeV range. Our results follow that trend because the absorption model presented here has many free parameters, and in particular, those related to the absorbing external fields are essentially unknown since the core of Cen A is so well enclosed by the host galaxy.\\

Following the literature we have considered a thin disc plus a hot corona to describe the emission of the accreting flow. But at the high accretion rate (Eddington) considered here, the actual case can be different, for example forming a thick disc as in Begelman \& Meier (1982). Then advective solutions might exist for the inflow and the absorption could be much lower that what we estimated. \\

Finally, if the strength of the magnetic field outside the jets is low enough, the development of IC electromagnetic cascades will be unavoidable given the great density of the radiation fields. Then the power contained in the photons with energy $\ga 10^{14}$ TeV that emerge from the jet can be reprocessed to lower energy photons, and the SED presented in Figure~\ref{sed} should be taken as a lower estimate for energies $E_\gamma\ga 1$ GeV. Future, longer exposures with HESS or HESS II could detect the source, with a rather soft spectrum above 200 GeV. \\

\medskip

{\em Acknowledgements:}
We are indebted to J.-P. Lenain for kindly sharing with us the observational points plotted with the SED and J.-P. Lenain, C. Boisson, A. Zech, and H. Sol for valuable discussions. This research has been supported by CONICET (PIP 5375) and the Argentine agency ANPCyT through Grant PICT 03-13291 BID 1728/OC-AR. G.E.R. is grateful to LUTh at Paris Observatory, for kind hospitality.

\label{lastpage}

\end{document}